# Multi-Tau Pulsed Illumination Differential Dynamic Microscopy with 80 µs Resolution

Emmanuel Schaub[1,a)], Martinus H. V. Werts[2]

[1]*Univ Rennes, CNRS, Institut de Physique de Rennes – UMR 6251, F-35000 Rennes, France*
[2]*Univ Angers, CNRS, MOLTECH-Anjou, SFR Matrix, F-49000, Angers, France*

[a)]Author to whom correspondence should be addressed: emmanuel.schaub@univ-rennes.fr.

**ABSTRACT**
*Differential dynamic microscopy (DDM) is a digital video-microscopy technique for quantitative measurements of the microscale dynamics in soft condensed matter systems. Here, multi-tau pulsed illumination DDM (MTPI-DDM) is introduced as a method for significantly enhancing the time resolution of DDM. The technique employs simple, low-cost instrumentation comprising a single monochrome digital camera and a single pulsed LED. A timing sequence, following a geometric progression of time lags, is used to generate a "multi-tau" scheme, providing high sampling density at short timescales where dynamics are fastest. In the current implementation, a time resolution of 80 µs is achieved, limited by the dead time of the camera electronics. Validation of MTPI-DDM was performed by measuring the diffusion of 147 nm polystyrene nanoparticles in water. Compared to conventional continuous-wave (CW) DDM, the pulsed approach extends the range of the shortest measurable time lags by nearly two orders of magnitude and enhances DDM signal amplitudes by eliminating motion blur.*

## I. INTRODUCTION

Differential Dynamic Microscopy (DDM) is a powerful and versatile modern optical microscopy technique for measuring dynamics at the micro-scale that has rapidly become very valuable in soft condensed matter physics[1,2]. Unlike single-particle tracking methods,[3,4] which require high spatial and temporal resolution and the elaborate identification and analysis of individual particle tracks, DDM extracts quantitative dynamic information from a sequence of standard, often low-resolution, optical microscopy images. Fundamentally, DDM operates by analyzing the temporal fluctuations of image intensity patterns by calculating the Fourier-space image structure function (ISF), also known as the differential image contrast function (DICF). This approach allows one to determine key parameters such as particle size distributions,[5] diffusion coefficients, including under flow conditions,[6] for a wide array of systems, from colloidal suspensions and bacterial cultures to polymer solutions.

The efficacy of DDM is fundamentally linked to its temporal resolution, related to the rate at which image frames are captured[1,7]. High temporal resolution is paramount because DDM measures dynamics by quantifying how image decorrelation unfolds over time. If the sampling rate is too slow, rapid motions are missed, leading to an incomplete or erroneous interpretation of the underlying physical processes. Conversely, an appropriately high frame rate captures the full spectrum of dynamics, enabling the precise measurement of fast diffusion and rapid flow velocities. Therefore, optimizing temporal resolution is critical to defining the limit of detectable motion, ensuring the fidelity of the extracted parameters, and ultimately unlocking the full potential of DDM to probe fast and complex dynamics *in situ*.

To improve time resolution, Arko and Petelin[8] developed a microscope setup equipped with two cameras for the study of the wavevector-dependent dynamics of soft matter with DDM. The cameras are triggered randomly acquiring two sequences of images of the same region in the sample. From the two data sets cross-image differences are calculated, which are subsequently analyzed in Fourier space as a function of time delay between the two images ('cross-DDM'). With this technique, an effective 125 µs time resolution was reached. This improved time resolution comes at the cost of elaborate precise alignment of the two cameras.

In the same vein, You and McGorty[9] developed a two-color DDM approach, using a single color camera. The method consists in sequentially illuminating the sample with spectrally distinct light pulses (red and blue). By pulsing the blue and then the red light separated by a time lag much smaller than the exposure time of the camera, dynamics much faster than the camera frame rate were captured, with an ultimate time resolution of 3 ms.

Mention should also be made of the analysis developed by Brogioli et al.,[10] based on the change in the power spectrum of microscope images as a function of the exposure time, which enables measurement of the fast dynamics of a colloidal suspension with high time resolution, not using DDM. However, this technique requires much time for data acquisition, because a large number of images is required for each exposure time with a sufficient delay between images so that they are fully decorrelated.

In the present work, we propose an alternative pulsed illumination method for DDM which only uses a single low-cost monochrome camera and a single pulsed LED for illumination. In its current configuration, our setup achieves a resolution of 80 µs. The proposed method consists in illuminating the sample with a short LED pulse, while imaging the sample at the maximum exposure time for the chosen frame-rate of the camera. With a single, specifically timed, LED pulse per frame, the time lag between two frames is thus fully determined by the time between two LED pulses. The minimum time lag is limited by the dead time of the camera and by the width of the LED pulses. This principle of pulsed illumination, which allows the minimal interval between two exposures to be limited only by the camera's dead time, is conventionally implemented in double-pulsed illumination PIV,[11] and here it is applied in DDM for the first time.

The LED pulse sequence is designed such that the time lags follow a geometrical law, giving equidistant $\tau$ values on a logarithmic scale, similar to that used in the multi-tau scheme known from fluorescence correlation spectroscopy and dynamic light scattering[12,13]. The limit for the longest accessible time lag is determined by the full acquisition time. The method is dubbed "multi-tau pulsed illumination DDM" (MTPI-DDM).

## II. PRINCIPLE

In DDM, the images of a video-microscopic sequence are analyzed in pairs, each pair corresponding to the time lag $\tau$ between the two images. The two-dimensional Fourier transforms of the differences between the images in the pairs are stored as a function of time lag $\tau$, giving the image structure function[14] (ISF, also called DICF: differential image correlation function[15]). For spatially isotropic dynamics, such as Brownian motion, the azimuthal average of the DICF is typically considered, $\mathrm{DICF}(q,\tau)$, where $q$ is the wavenumber.

The intensity of the DICF is plotted and analyzed as a function of $\tau$, for various values of $q$. In many cases, notably in the study of diffusion processes, the plots are represented on a logarithmic timescale, in order to visualize the phenomena over a wide time range. Having a sequence of equidistant $\tau$ values in a logarithmic scale, means that the series of $\tau$ values must follow a geometrical progression. Thus, a set of images is desired whose time differences are distributed between a minimum and a maximum according to a geometric law:

$$\tau_1 = \tau_{\min}$$
$$\tau_{i+1} = \alpha\tau_i$$
$$\tau_{\max} = \tau_{N_1} = \alpha^{N_1-1}\tau_{\min}$$

In the conventional DDM approach, under continuous wave illumination, the time difference between two successive images is constant and defined by the camera frame rate. In pulsed illumination, with one single pulse per exposed image, the effective time lag between two images is the time difference separating the two light pulses. This time difference can be tuned between a minimum value limited by the dead time of the camera to larger values limited by the exposure time.

Furthermore, pulsed illumination allows the sample to be imaged over a specific, very short time period, whereas constant intensity illumination results in an image integrated over the exposure time. This results in blurring in the case of rapidly moving objects, which affects the DICF.[16]

### A. Illumination scheme

In MTPI-DDM, a series of digital video frames is acquired with a single camera at a constant frame rate, while applying a specific pulsed illumination sequence to obtain the multi-tau sequence with high time resolution (Fig. 1).

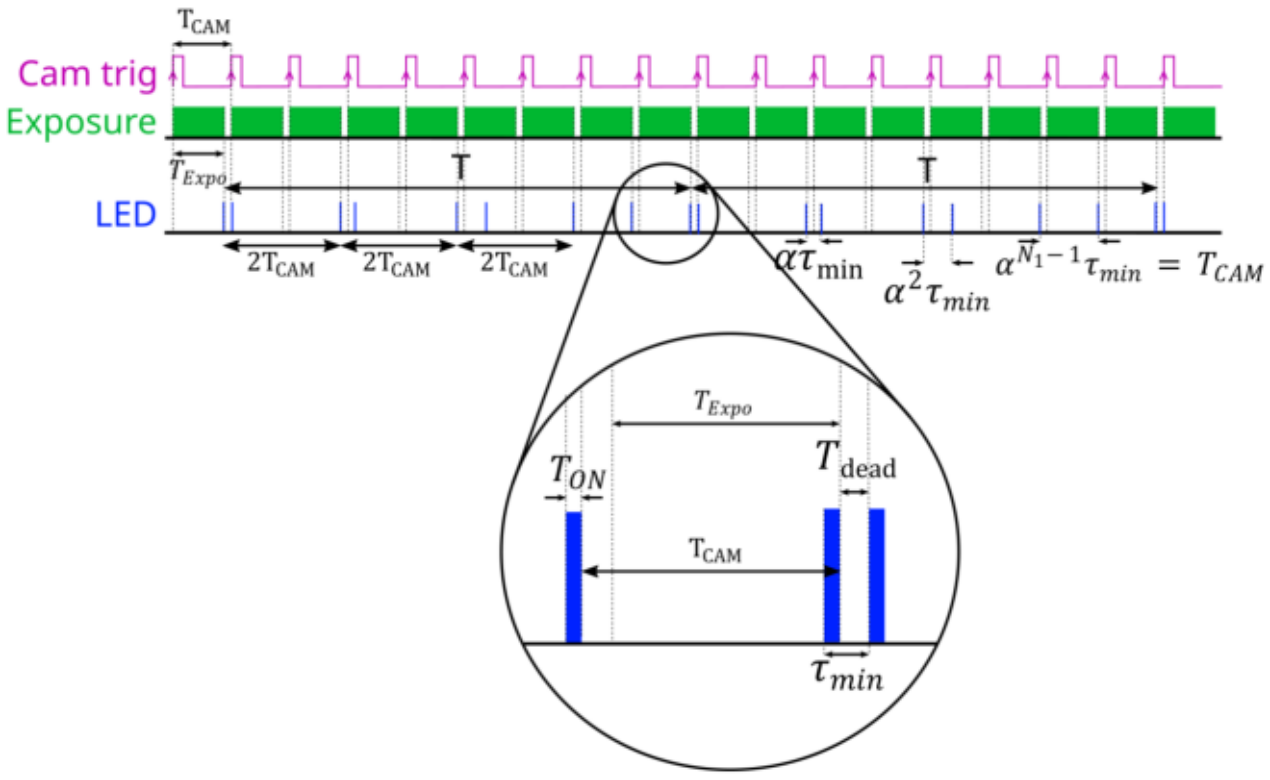

FIG. 1. MTPI-DDM Illumination and acquisition scheme. (pink): Camera trigger signal, triggering the image acquisition on rising edges. (green): Camera exposure, set to maximum available exposure time. The camera exposures are separated by the camera dead time $T_{\mathrm{dead}}$. (blue): LED pulses. Exactly one pulse for each exposure time. $T$ is the pulse train period.

The upper signal in Fig. 1 is the trigger signal of the camera. It has steady period $T_{\mathrm{CAM}}$, defined by the frame rate of the camera. The second signal represents the frame exposure periods. It has the same frequency as the trigger signal. The exposure time $T_{\mathrm{Expo}}$ is slightly shorter than the camera period $T_{\mathrm{CAM}}$, because of the camera dead time between two successive images: $T_{\text{dead}} = T_{\text{CAM}} - T_{\text{Expo}}$. For the camera used in the set-up, this dead time is 60 µs. Exactly one, and only one, LED pulse is fired during the exposure time of each frame. The width of the LED pulses is $T_{ON}$ . In the present case, $T_{ON} = 20$ µs.

Trains of $N_1$ pairs of pulses are generated periodically. The full period for these trains is $T = 2N_1T_{\text{CAM}}$. The pulse time is programmed as follows:

- The first LED pulse of the pulse train arrives at the end of the exposure time of the first image, at time $t_1 = T_{\text{Expo}} - T_{\text{ON}}$.
- The second LED pulse arrives at the very beginning of the exposure of the second image, at time $t_2 = T_{\text{CAM}}$.
- The third LED pulse arrives at the end of the exposure time of the third image, at time $t_3 = \text{t}_1 + 2\,T_{\text{CAM}}$
- The fourth LED pulse arrives after $\alpha\tau_{\text{min}}$, at time $t_4 = t_3 + \alpha\tau_{\text{min}}$.

The LED illumination pulses are thus timed according to the following equations:

$$t_{2i+1} = t_{2\text{i}-1} + 2T_{\text{CAM}}, \text{ for } i = 1, 2, \dots, N_1 - 1$$
$$t_{2\text{i}+2} = t_{2i+1} + \alpha^i\tau_{\text{min}}$$

$\alpha$ is chosen precisely such that the last two pulses of the train are separated by $T_{\text{Cam}}$, which leads to:

$$\alpha = \sqrt[N_1-1]{T_{\text{CAM}}/\tau_{\text{min}}}.$$

Thus, the pulses corresponding to the odd-indexed images are uniformly separated by a time $2T_{\text{CAM}}$.

The ultimate time resolution $\tau_1 = \tau_{\text{min}} = t_2 - t_1 = T_{\text{dead}} + T_{\text{ON}}$ is 80µs. It is limited by the dead time of the camera $T_{\text{dead}}$, and $T_{\text{ON}}$ the pulse duration.

### B. Image sequence analysis

As mentioned above, in DDM, images from the video sequence are analyzed in pairs to obtain the DICF, ideally with a time lag between images in subsequent pairs that follows a geometric progression with a common ratio $\alpha$. The analysis of the MTPI-DDM image sequences consists of two main parts, one giving the short-time DICF, the other giving the long-time DICF.

The short-time DICF is obtained from analysis of the first $N_1$ values of the geometric progression of $\tau$, at time lags $\tau_i = \alpha^{i-1}\tau_{min}$, with $1 \le i \le N_1$, only processing image pairs with indices $2i - 1$ and $2i$ within a pulse train. The results are averaged over the same time lags in all subsequent pulse trains in the complete image stack, giving the DICF for the shorter time lags.

For DICF at long times, odd-indexed images correspond to evenly spaced sampling in time, with a period $2T_{\text{CAM}}$. It is thus possible to compute the DDM for all values of $\tau$ that are multiples of $2T_{\text{CAM}}$ giving the DICF for the longer time lags. This part of the analysis is done applying the Wiener-Khinchin theorem, as described by Norouzisadeh et al.[17] Subsequently, we reduce the number points by averaging the DICF around $\tau$ values that follow the desired geometric progression.

The initial points of the long-time DICF curve are quite spaced out and do not follow the desired geometric progression. In order to ensure a good overlap between the short- and long-time DICF and continuity in the geometric progression of time lags, additional pairs of images were chosen in each pulse train to obtain the DICF at intermediate time lags. Good continuity between the short-time and long-time DICF was found, and no adjustment was required between the long- and short-time parts (Suppl. Mater. A).

## III. EXPERIMENTAL SETUP

A standard commercial Olympus IX71 inverted microscope is used in transmitted bright-field configuration, with a x20 objective lens (Olympus, UPlanSApo 20x/0.75). A condenser with a numerical aperture of 0.55 was used, but its aperture was minimized to enhance both contrast and depth of field. The sample is imaged with a USB3 CMOS camera (Basler ACE-acA1920-155um), capable of acquiring images at 165 frames per second (fps) full frame (1920x1200 pixels).

Given that the LED pulses need to be as short as possible, they have to be intense in order to provide sufficient radiant energy to the image sensor for each image. In designing our LED driver, we were inspired by the work of Willert *et al.*[18] The schematic of the driver is shown in Fig. 2. A National Instruments module (NI-USB 6343) generates three electronic control signals. Output signal Y1 indicates the start of the pulse trains. It is used to trigger an oscilloscope for visualizing the other signals. Output Y2 is the camera trigger signal. Output Y3 is the LED pulse gate signal. The analog input signal Y4 is used to monitor the current through the LED via shunt resistance $R_s$. Output signals Y1, Y2 and Y3 are generated using analog outputs of the NI-USB 6343 module. The amplitude of Y1 and Y2 is 5V. The amplitude of Y3 is 10V.

The MOSFET transistor Q operates in switching mode. The current through the LED is controlled by the regulated power supply voltage $V_{DD}$ (approx. 6V). The current is monitored on the oscilloscope via the Y4 signal. We use a high-power green LED (OSRAM, LST1-01F06-GRN1-00). With this LED, the current pulses have a width of 20µs and an amplitude of 20A. Surprisingly, although designed to withstand a maximum CW current of 700mA, the LED can withstand current as high as 20A during microsecond pulses. A high current in the diode is all the more important as the microscope condenser aperture is reduced, resulting in less light illuminating the sample.

Under these operating conditions, respecting short current pulse durations, no LED damage occurred and reliable image acquisition could be performed indefinitely. However, longer pulses, even of only a few millisecond duration, will lead to violent and irreversible LED damage.

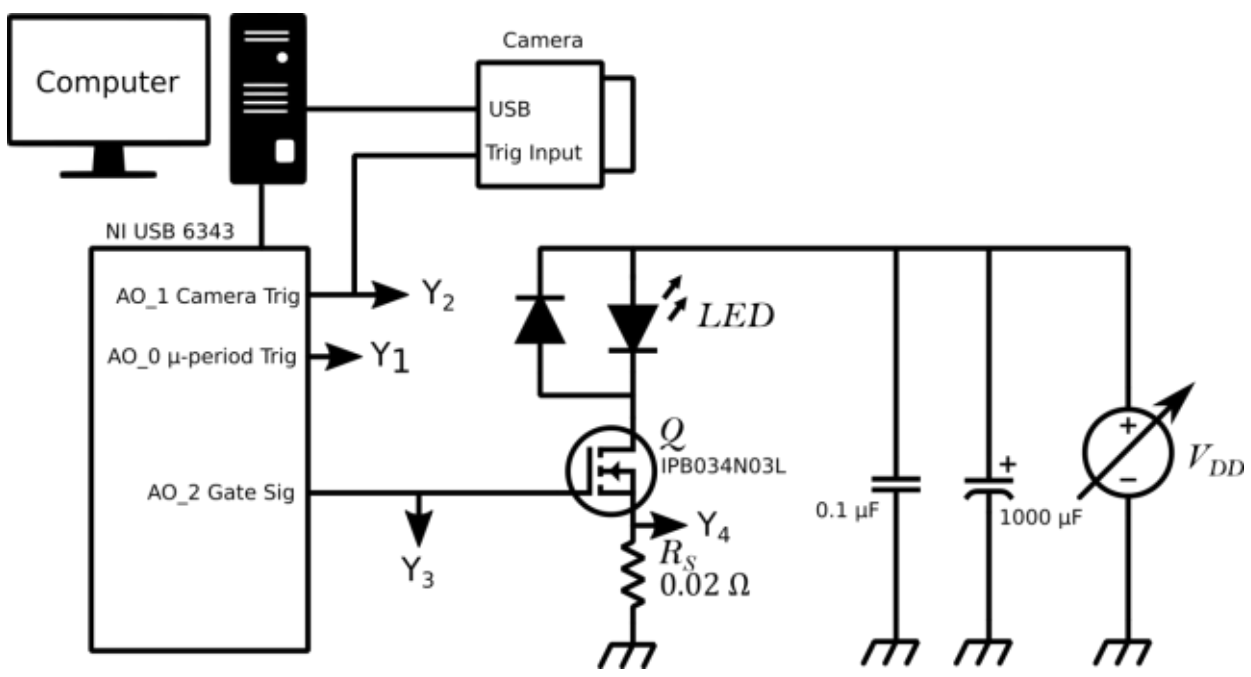

FIG. 2. Driver circuit for pulsed LED operation with synchronized image acquisition.

## IV. RESULTS AND DISCUSSION

MTPI-DDM measurements were performed using a reference sample of polystyrene (PS) nanospheres in aqueous solution, with a specified diameter of (147 ± 4) nm (PS Research, Germany). Measurements were carried out at two concentrations: 0.5 g/l and 0.05 g/l, obtained by diluting the stock solution with filtered ultrapure water. Each acquisition consisted of a series of 2000 images with a resolution of 1200 × 1200 pixels and 8-bit depth. The frame rate was set to 165 fps, corresponding to an acquisition time of 12.1 s. Each acquisition produced an image stack of approximately 2.8 GB. Measurements were conducted for $N_1$ = 10, 20, 30, and 40, and each condition was repeated 10 times. The DDM analysis was carried out in Python, utilizing the NumPy and SciPy libraries for numerical processing and Matplotlib for visualization.

Figure 3 shows the DICF curves obtained, on one hand, under CW illumination and, on the other hand, using the presently described MTPI-DDM, for $N_1$ = 20, and a NP concentration of 0.5 g/l. Data obtained under pulsed and CW illumination are shown as blue and red curves, respectively, for a range of wavenumbers $q$. The range includes smaller values of $q$, for which the characteristic relaxation time is sufficiently long and the dynamics are within the accessible $\tau$ range for CW-DDM, allowing the CW and MTPI curves to be compared directly. The acquisitions performed in CW and MTPI were carried out using the same camera settings (frame rate, gain, exposure time). To account for the illumination difference, the CW DICF was normalized with respect to the MTPI DICF. The normalization factor was determined as the ratio of the MTPI DICF amplitude to that of the CW DICF at $q$ = 2 µm⁻¹, corresponding to a low wavevector for which the exposure time does not influence the CW DICF amplitude (Suppl. Mater. B).

The CW-DDM curves start at a minimum time lag of 6.1 ms, which corresponds to $T_{\mathrm{CAM}}$, the frame period of the camera. For pulsed illumination, the curves begin at a minimal time lag of 80 µs, corresponding to the minimum applicable time between two pulses, which is limited by the camera deadtime. We have thus improved the temporal resolution by nearly two orders of magnitude. We also observe that, while for low values of $q$ the CW and MTPI curves superimpose, at higher $q$ values the amplitude of the CW curve is reduced compared to that of the MTPI curves. In the case of Brownian diffusion of non-interacting nanoparticles (NPs), it is well known that the amplitude of the DICF depends on the microscope setup, the particle form factor, and their concentration.[19] Here, we show that at high wavevector values, this amplitude is also affected by the image exposure time.

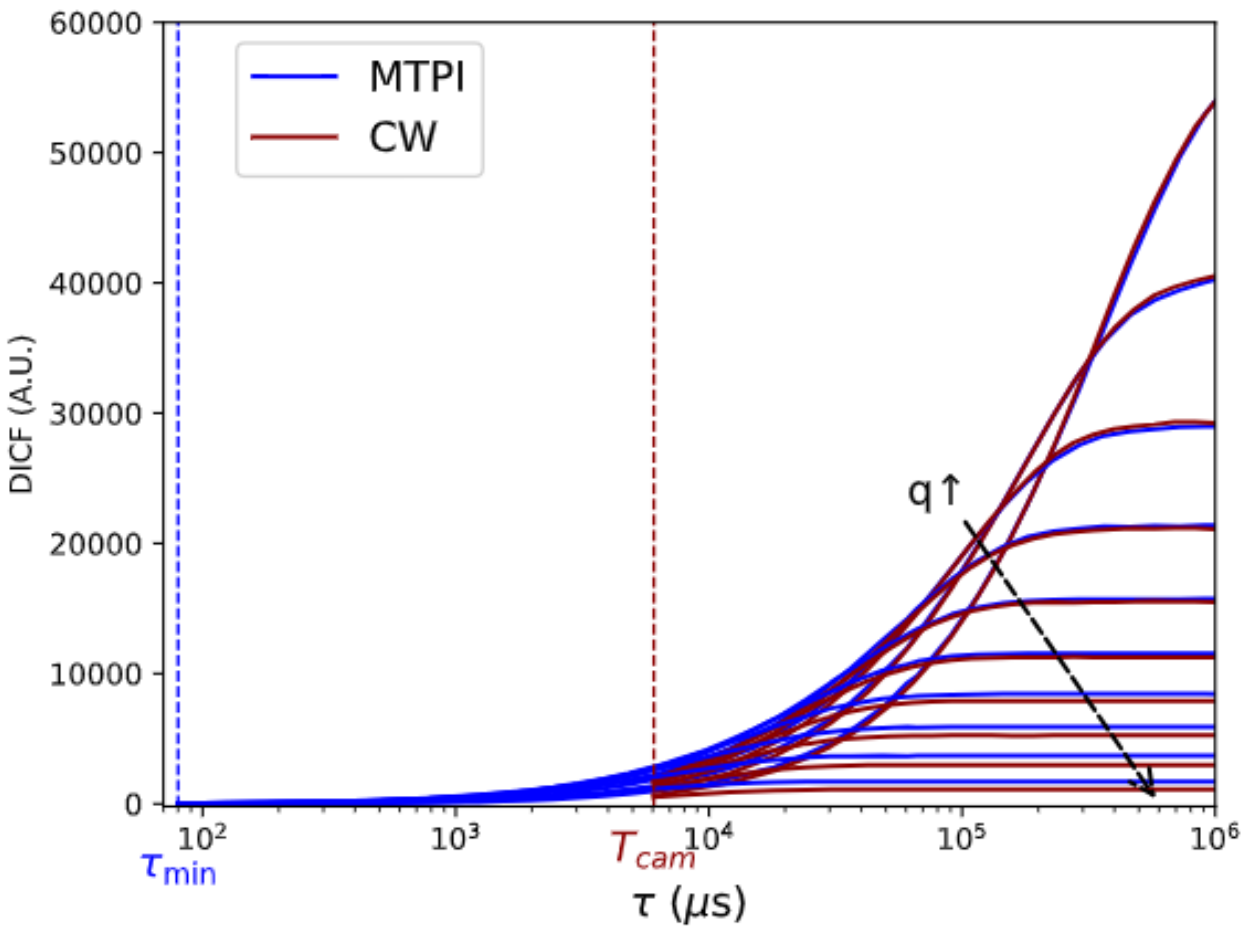

FIG. 3: Comparison of the azimuthally averaged DICFs (147 nm PS NPs, 0.5 g/l in water at 19°C), between CW illumination (red curves) and pulsed illumination (blue curves), for q = 1.00, 1.45, 1.97, 2.56, 3.23, 4.00, 4.84, 5.75, 6.75, 7.85 µm$^{-1}$

Zooming in on the differences between the CW and MTPI DICFs, Fig. 4a presents DICF curves for four $q$ values ranging from 4.84 µm⁻¹ to 7.85 µm⁻¹. For these large $q$ values, the amplitude of the CW curves is clearly lower than that of the MTPI curves. A further zoom (Fig. 4b) demonstrates that the shortest time lags $\tau$, inaccessible to CW-DDM, the MTPI DICF curves are still very well suited for analysis, even though they appear almost flat at short times in Fig. 4a.

In CW-DDM, motion blur occurs due to the long image exposure time, significantly reducing the amplitude of the DICF. This effect can be quantified in the form of a blurring factor, defined as the ratio of the CW DICF amplitude to that of the MTPI DICF (Suppl. Mater. B). In Fig. 4c, this blurring factor is plotted. The blurring becomes increasingly pronounced at increasing $q$.[16] Derivation of an analytical expression for this blurring factor might be feasible, but would be beyond the scope of this paper. The critical wave vector beyond which the blurring effect becomes significant is expected to be on the order of $1/\sqrt{T_{\mathrm{CAM}}D} \approx$ 7.5 $\mu\mathrm{m}^{-1}$ for 147 nm NPs in water at 19°C, where the diffusion coefficient $D = 2.9\ \mu\mathrm{m}^2/s$.

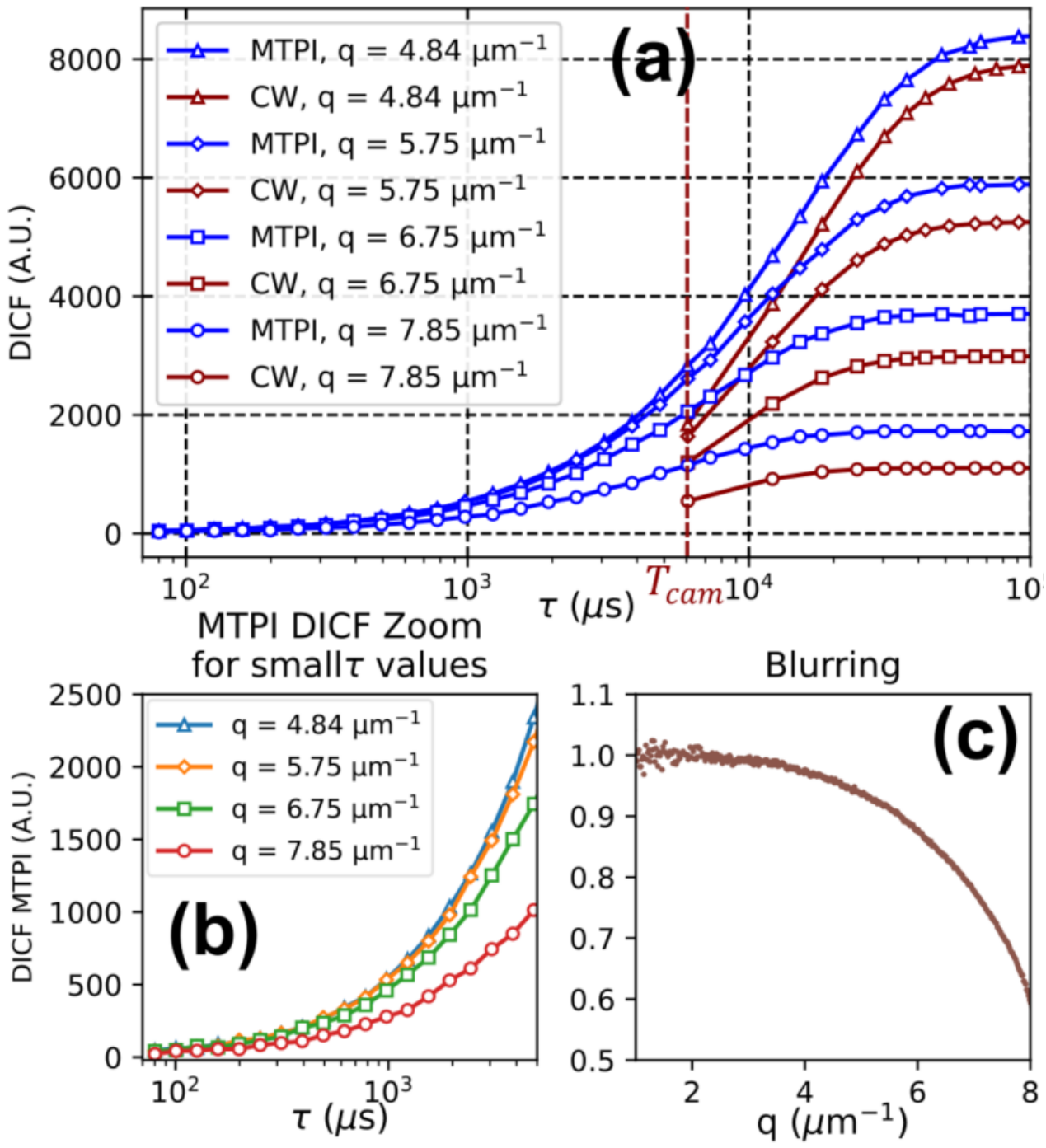

FIG. 4. (a) Comparison of the azimuthally averaged DICFs obtained with CW illumination (red curves) and pulsed illumination (blue curves), for wavenumbers ranging from 4.84 to 7.85 µm⁻¹, highlighting the extension of the accessible wavenumber range towards lower $\tau$ values (down to 80 µs) enabled by pulsed illumination. (b) Zoom at short τ. (c) Blurring factor, defined as the ratio of the CW DICF amplitude to the MTPI DICF amplitude.

The DICF curves were analyzed by curve-fitting the classical model for DDM[20]

$$\mathrm{DICF}(q,\tau) = A(q)\big(1 - f(q,\tau)\big) + B(q) \qquad (1)$$

The function $f$ is identified with the intermediate scattering function (ISF). For a solution of freely diffusing monodisperse NPs in 2D imaging, $f(q,\tau) = \exp(-\tau/\tau_c)$, with time constant $\tau_c = 1/q^2 D$. $D$ is the diffusion coefficient, which is related to the hydrodynamic diameter $d_h$ of the NP through the Stokes-Einstein-Sutherland equation

$$d_h = \frac{kT}{3\pi\eta(T)D} \qquad (2)$$

In (2), $k_B$ is the Boltzmann constant, $T$ is the temperature, $\eta(T)$ is the viscosity of the liquid medium at that temperature (for water, see Ref. 21), and $d_h$ the hydrodynamic diameter of the particle.

The DICF was fit independently for each value of $q$ using Eq. 1, giving a characteristic time constant $\tau_c$ for each $q$ (Suppl. Matter. A). In Figure 5, the corresponding diffusivity $D = 1/\tau_c q^2$ is plotted as a function of $q$. This was done for the DICF obtained by CW-DDM (red) as well as for the DICF obtained by MTPI-DDM (blue).

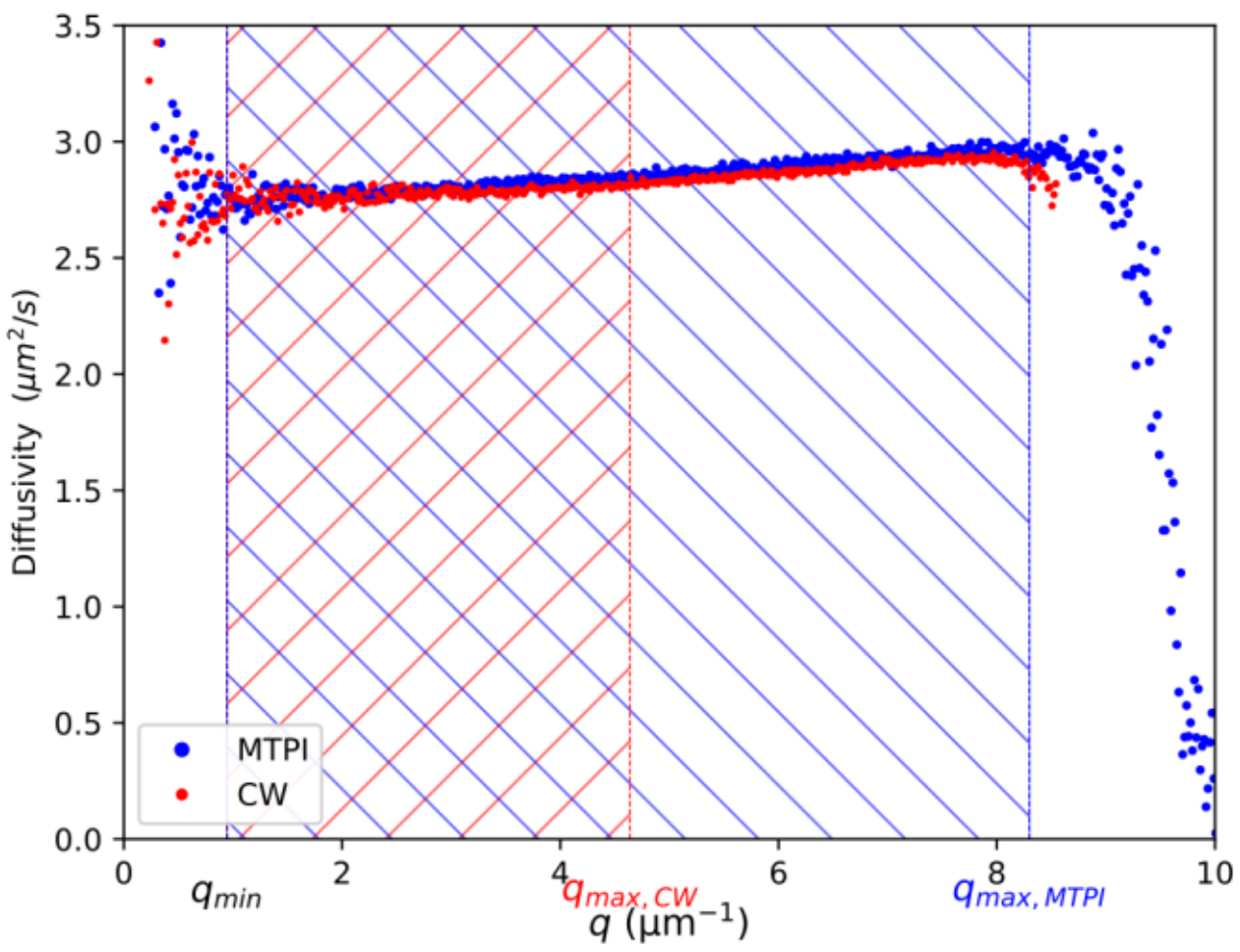

FIG. 5. Diffusivity as a function of the wavenumber $q$ for CW illumination (red) and MTPI-DDM (blue). $q_{\min}$ is the minimum wavenumber value used to estimate the mean diffusivity. $q_{\max,CW}$ is the maximum wavenumber value used to estimate the mean diffusivity in CW illumination. $q_{\max,MTPI}$ is the maximum wavevector value used to estimate the mean diffusivity in MTPI-DDM (see text).

In the central portion of the plot, both the CW and MTPI curves are practically superimposed. Both curves are approximately constant, only very slightly increasing with increasing $q$. At very low and very high $q$ values, the data becomes noisy and cannot be analyzed.

A reasonable unbiased criterion needs to be established for deciding which range of $q$ will be used to estimate the diffusivity as the mean value over that range. The time constant $\tau_c$ for each $q$ corresponds to the inflection point of the DICF curve when plotted on a logarithmic time axis, such as in Figure 4a. To reliably determine $\tau_c$, and therefore the diffusion coefficient $D$, it is necessary to consider a sufficiently wide portion of the curve around this inflection point that allows identifying the actual change in concavity rather than local fluctuations due to noise. A simple reasonable criterion is that the DICF curve for a given $q$ should contain the extrema of the second derivative (of the curve on the logarithmic time scale). For our ISF model $f(q,\tau) = \exp(-\tau/\tau_c)$, the third derivative (on the logarithmic time scale) vanishes at $\tau_1 = \tau_c \times (3-\sqrt{5})/2 \approx 0.38\,\tau_c$ and $\tau_2 = \tau_c \times (3+\sqrt{5})/2 \approx 2.6\,\tau_c$. This means that the experimental curve should encompass data points between $0.38\,\tau_c$ and $2.6\,\tau_c$.

The upper limit $\tau_2$ is set by the acquisition time and by the fact that the last few points, corresponding to the highest

$\tau$ values, are noisy. In practice, the points on the DICF curves were discarded for $\tau$ values greater than $\tau_{\max} = 1$ s. This constrains the wavevector to a minimal value, $q_{\min} = \sqrt{2.6/(\tau_{\max} D)}$. For 147 nm NPs in water at 19°C, this corresponds to $q_{\min} = 0.94\ \mu\text{m}^{-1}$.

The lower limit $\tau_1$ imposes an upper bound on the $q$ values for which the characteristic time of the DICF can be reliably extracted. For CW illumination, this is $q_{max,CW} = \sqrt{0.38/(T_{\text{CAM}} D)} = 4.6\ \mu\text{m}^{-1}$. For MTPI, we find $q_{\text{max,MTPI}} = \sqrt{0.38/(\tau_{\min} D)} = 40\ \mu\text{m}^{-1}$. This latter value cannot be reached, because of the optical resolution limit of the microscope, which gives $q_{\text{max,diffr}} = 2\pi NA/\lambda_0$, in this case amounting to 8.3 $\mu\text{m}^{-1}$. Thus, in MTPI-DDM, the usable $q$ range is limited only by optical diffraction. This holds as long as the NP diameter is larger than $d = \frac{4\pi NA^2 k_B T \tau_{\min}}{0.38\times 3\eta\lambda_0^2} = 6.0$ nm. For NPs smaller than 6 nm (in water at 19°C), $q$ is limited by $\tau_{\min}$.

Finally, for our 147 nm NPs, under CW illumination, we estimated the diffusivity as the average of $D = 1/\tau_c q^2$ for $q$ values between $q_{\min} = 0.94\ \mu m^{-1}$ and $q_{\text{max,CW}} = 4.6\ \mu\text{m}^{-1}$. Under MTPI, we take the average between $q_{\min} = 0.94\ \mu\text{m}^{-1}$ and $q_{\text{max,diffr}} = 8.3\ \mu\text{m}^{-1}$.
The fact that the diffusion coefficient trace obtained by CW-DDM coincides with that obtained by MTPI-DDM, even for $q$ values higher than $q_{max,CW}$ would suggest that our criterion for choosing the $q$ range may be overly conservative. A more in-depth investigation to establish a refined mathematical criterion, however, lies beyond the scope of the present article.

The hydrodynamic diameters obtained are shown in Table I, both for CW-DDM and MTPI-DDM, at two different concentrations of 147 nm PS NPs. Measurements in MTPI-DDM were performed for four values of $N_1$: 10, 20, 30, and 40. For each set of conditions, the measurements were repeated 10 times. The uncertainty is expressed as the 95% confidence interval over these 10 measurements.

TABLE I. Measured diameters (nm, reported with 95% confidence interval) of 147 nm PS reference nanoparticles in water at two concentrations, using MTPI-DDM, for $N_1$ parameter values ranging from 10 to 40, and using CW-DDM.

| conc. g/l | $d_{\text{spec}}$ nm | MTPI diam. (nm) | | | | CW diam. (nm) |
|---|---|---|---|---|---|---|
| | | $N_1$ | | | | |
| | | 10 | 20 | 30 | 40 | |
| 0.5 | 147 | 144.7 ± 0.1 | 144.2 ±0.1 | 144.5 ±0.1 | 144.2 ±0.1 | 141.4 ±0.1 |
| 0.05 | | 149.2 ±1.2 | 149.1 ±0.2 | 149.5 ±0.2 | 150.2 ±0.2 | 140.2 ±0.2 |

The hydrodynamic diameters obtained by MTPI-DDM are in close agreement with those from CW-DDM. Both agree with the manufacturer-specified diameter $d_{\text{spec}} = 147$ nm. Comparison of MTPI and CW shows that the exposure time in DDM affects the amplitude of the DICF but does not alter the functional dependence of the relaxation on $\tau$. The standard deviations indicate that the measurements are highly reproducible. The measurement does not depend on the value of the parameter $N_1$, at least within the explored range of 10…40.

## V. CONCLUSION

We developed a pulsed illumination device using a single pulsed LED and a single camera. It enables MTPI-DDM measurements with high temporal resolution, essentially limited by the camera dead time. A multi-tau acquisition scheme was implemented that is well suited to the study of diffusive processes and other dynamics in soft condensed matter physics, taking fewer data points on the time scale where the DICF changes slowly and more points where it changes rapidly. The advantages of the MTPI scheme compared to standard CW illumination are two-fold. Firstly, the accessible time range is extended towards very short time scales, allowing reliable curve-fitting for analysis for higher wavenumbers. Secondly, the amplitude of the DICF is increased as a result of the suppression of motion blur.

A limitation of MTPI-DDM is that it requires the use of a high-intensity pulsed LED source and synchronization electronics, which are not commercially available. Another potential limitation arises from the fact that only one image pair per train is used to compute the short-time DICF points, in contrast to the long-time DICF points (Wiener–Khinchin method). Consequently, for very dilute samples, the short-time DICF points may exhibit higher noise levels. In such cases, a longer acquisition time would be required. However, within the scope of this work, this did not constitute a limitation, as demonstrated by the high reproducibility of the measurements.

MTPI-DDM is particularly valuable for accessing the high wavenumber (high $q$) regime in DDM. The increased time resolution and higher signal amplitude at high wavenumbers enable the measurement of faster dynamic processes, such as the short-time diffusion of smaller nanoparticles. More importantly, we expect that having access to both the high and low plateaus of the DICF on a logarithmic time-scale will be all the more critical when having a higher number of model parameters, for instance when analyzing polydisperse nanoparticle mixtures. More generally, the high temporal resolution of our apparatus will be useful for analyzing phenomena giving rise to rapid fluctuations, such as rotational diffusion[22] and liquid crystals fluctuations.[23]

### Supplementary material

Supplementary Material A details how the points for the short-time DICF are calculated and demonstrates that there is no discontinuity between the short-time DICF and the long-time DICF. Supplementary Material B presents the results of fits performed using Eq. (1) and explains how the DICF curves obtained by CW-DDM were rescaled relative to those obtained by MTPI-DDM, accounting for the difference in illumination.

### Conflict of Interest Statement

Emmanuel Schaub has patent FR2301561 pending.

### Data Availability Statement

The data that support the findings of this study are available from the corresponding author upon reasonable request.

# Supplementary Materials

# Multi-Tau Pulsed Illumination Differential Dynamic Microscopy with 80 µs Resolution

Emmanuel Schaub[1,a)], Martinus H. V. Werts[2]

*[1]Univ Rennes, CNRS, Institut de Physique de Rennes – UMR 6251, F-35000 Rennes, France*
*[2]Univ Angers, CNRS, MOLTECH-Anjou, SFR Matrix, F-49000, Angers, France*

[a)]Author to whom correspondence should be addressed: emmanuel.schaub@univ-rennes.fr.

## A. Short-time DICF

The following table presents the pairs of images within a pulse train that are used to calculate the short-time DICF, for $N_1 = 20$. The first column indicates the index of the first image (the index of the first image of the train is zero in this table). The second column indicates the index of the second image of the pair. The third column gives the time separating the two LED pulses of these two images.

The first $N_1 = 20$ image pairs follow the short-time progression of time lags. The additional pairs, indicated in italics, are selected from the same stack of 40 recorded images to cover time lags that create better overlap with the long-time (Wiener-Khinchin) DICF and create a better geometric progression of time lags. The initial points of the latter (long-time) DICF are quite spaced out and do not follow the desired progression.

| **Idx1** | **Idx2** | **time lag $\tau$ (µs)** |
|---|---|---|
| 0 | 1 | 80.0 |
| 2 | 3 | 100.46357632314935 |
| 4 | 5 | 126.16162709546569 |
| 6 | 7 | 158.4331031594753 |
| 8 | 9 | 198.95945189219177 |
| 10 | 11 | 249.85222600478963 |
| 12 | 13 | 313.7631022092618 |
| 14 | 15 | 394.0220420773535 |
| 16 | 17 | 494.81079371551715 |
| 18 | 19 | 621.3807742494621 |
| 20 | 21 | 780.3266854943558 |
| 22 | 23 | 979.9301190644047 |
| 24 | 25 | 1230.5910538497458 |
| 26 | 27 | 1545.369728262734 |
| 28 | 29 | 1940.667120535097 |
| 30 | 31 | 2437.079492271302 |
| 32 | 33 | 3060.4715197172504 |
| 34 | 35 | 3843.323926324236 |
| 36 | 37 | 4826.425832585761 |
| 38 | 39 | 6061.0 |
| | | |
| *37* | *38* | *7295.574167414239* |
| *31* | *32* | *9684.920507728697* |

(continued from previous page)

| *30* | *33* | *15182.47151971725* |
| --- | --- | --- |
| *36* | *39* | *18183.0* |
| *34* | *39* | *30305.0* |
| *5* | *15* | *60877.860414981886* |

*Table description: The table above consists of three columns. The first column shows the index of the first image in the image pair in question, within the pulse train. The second column shows the index of the second image in the image pair in question. The third column shows the time, in microseconds, between the two images in the image pair in question.*

Fig. S1 demonstrates the continuity between the short-time and long-time DICF. No scaling or adjustment between the short-time and long-time DICF was applied.

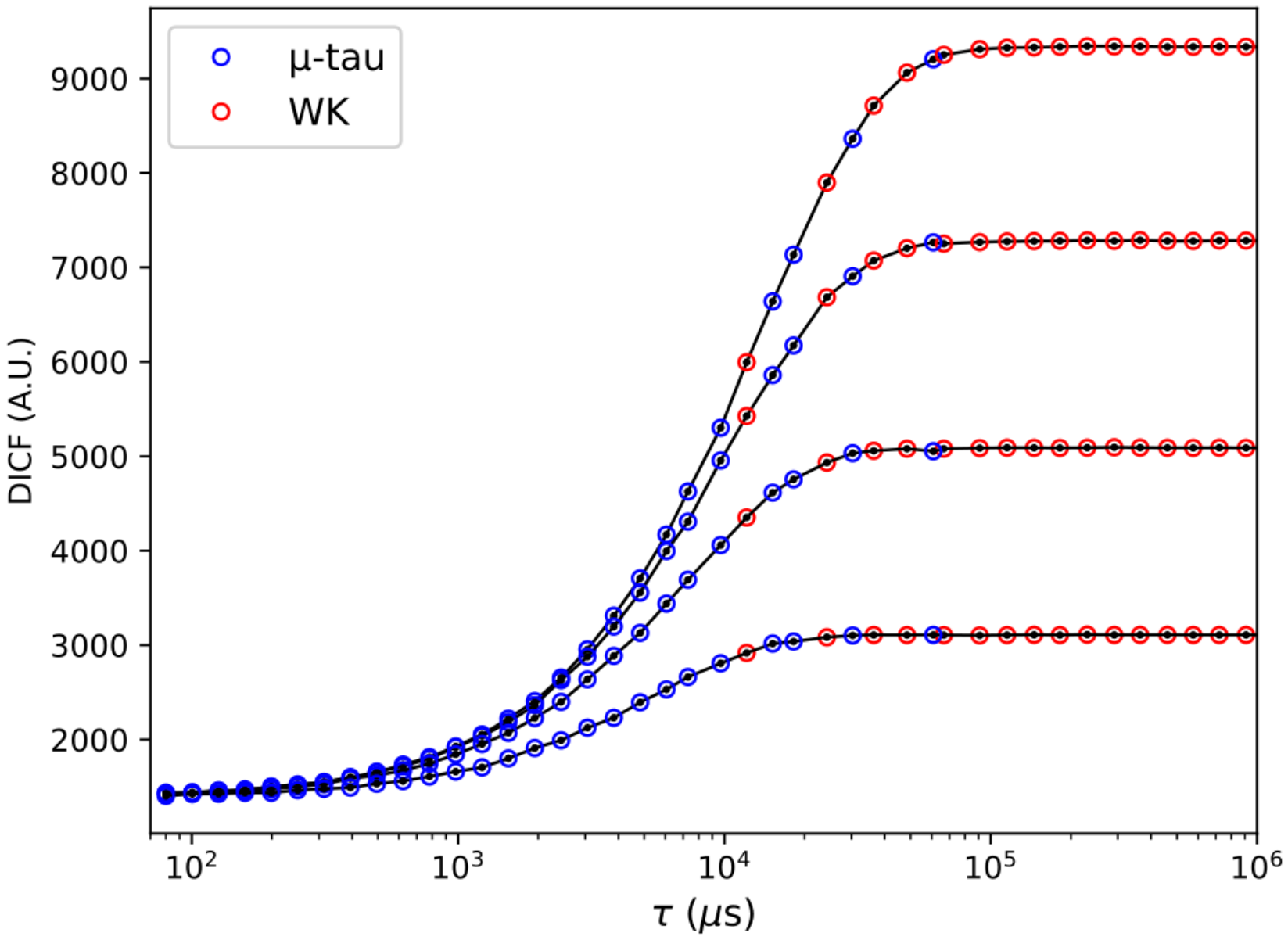

**Fig. S1.** DICF curves measured on 147 nm PS nanoparticles (0.5 g/l), for *q* values of 4.84, 5.75, 6.75 and 7.85 $\mu m^{-1}$. The blue points (“µ-tau”) correspond to the short-time DICF, and the red points (“WK”) to long-time (Wiener-Khinchin) DICF. This curve illustrates the continuity between the short-time and long-time DICF.

*Figure description: Figure S1 shows DICF curves obtained under pulsed illumination. The x-axis shows τ on a logarithmic scale, for values ranging from 80 µs to 1 s. The y-axis shows the DICF. Four curves are shown for values of q ranging from 4.84 to 7.85 $\mu m^{-1}$. These curves demonstrate that there is no discontinuity between the curve points calculated using the direct method and those calculated using the Wiener-Khinchin method.*

**B. Rescaling of CW DICF curves relative to MTPI DICF curves**

The CW and MTPI DICF curves were acquired using the same camera settings but under different illumination modes. In order to plot both curves on the same graph and allow for a fair comparison, CW DICF curves need to be corrected and rescaled.

To this end, curves were fit to the raw DICF data (CW and MTPI) for a range of wavenumbers $q$, according to the conventional model:

$$\mathrm{DICF}(\tau, q) = A_{\mathrm{fit}}(q)(1 - \exp(-\tau/\tau_{\mathrm{fit}}(q)) + B_{fit}(q)$$

In the curve fitting, all parameters $A_{\mathrm{fit}}(q)$, $B_{fit}(q)$, $\tau_{\mathrm{fit}}(q)$ were used as free adjustable parameters. The fit parameters obtained for the 0.5 g/l 147nm PS nanoparticle solution are plotted in Fig. S2.

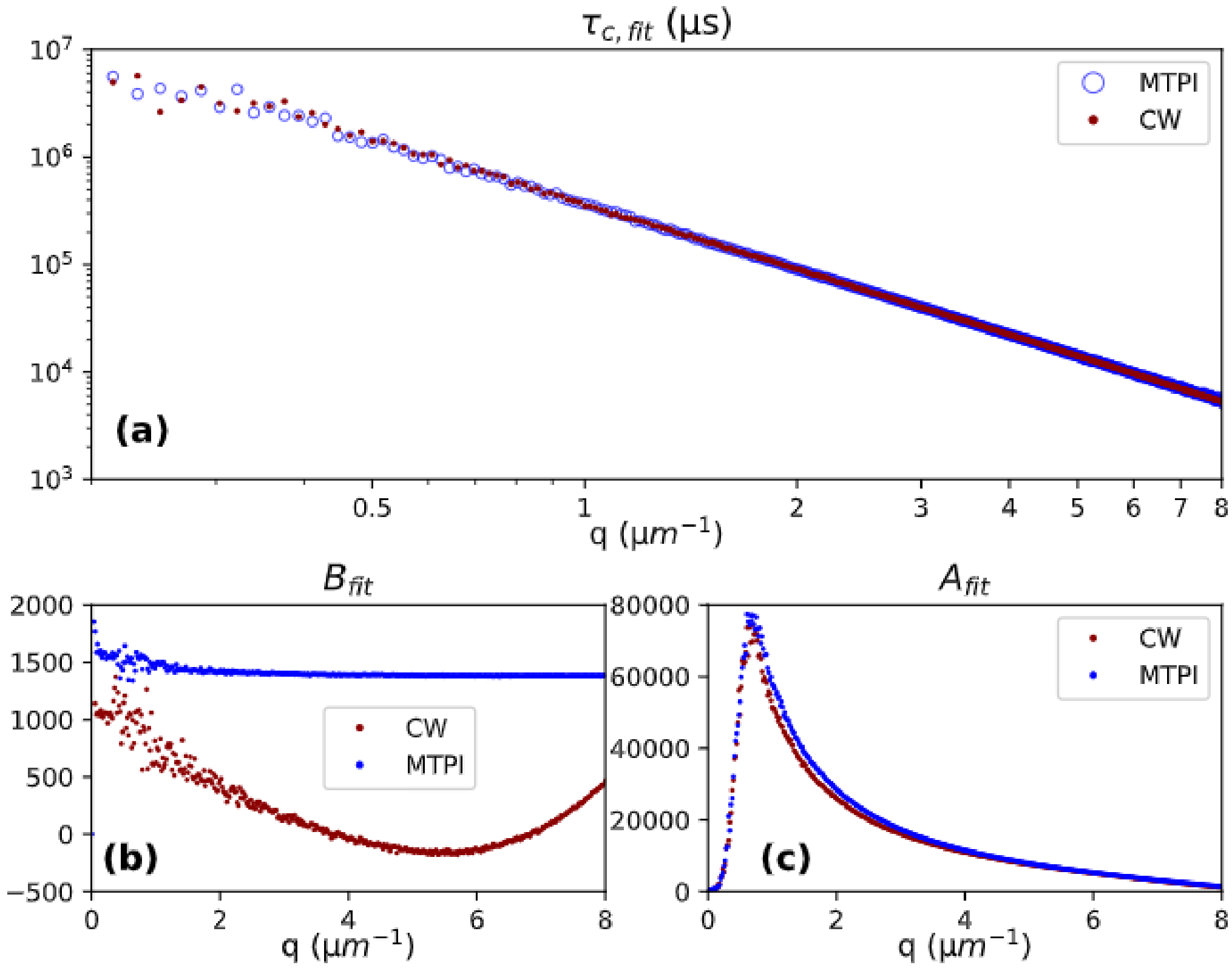

**Fig. S2.** Fit parameters obtained with all parameters left freely adjustable. (a) Characteristic time $\tau_{c,\mathrm{fit}}$. (b) $B_{\mathrm{fit}}$: offset (attributed, in principle, to electronic noise). (c) $A_{\mathrm{fit}}$: amplitude of the DICF.

*Figure description: Figure S2 shows the results of the DICF curve fits for acquisitions under CW and pulsed illumination. The figure consists of three panels. The top panel, a, shows the characteristic time as a function of q. The scale for q is logarithmic. The vertical scale for the characteristic time is also logarithmic. In this plot, the characteristic time data points obtained under pulsed illumination overlap with those obtained under CW illumination. Below on the left, panel b shows the parameter $B_{fit}$ as a function of q. It is constant in MTPI-DDM, whereas it is not under CW illumination. At the bottom right, panel c shows the estimated amplitude of the DICF, $A_{fit}$, as a function of q, for CW illumination and for pulsed illumination. The two curves are similar.*

The MTPI curves were corrected by simply subtracting the offset $B_{\mathrm{fit,MTPI}}(q)$:

$$\mathrm{DICF}_{\mathrm{MTPI,corrected}}(\tau,q) = \mathrm{DICF}_{\mathrm{MTPI}}(\tau,q) - B_{\mathrm{fit,MTPI}}(q)$$

The correction of the CW curves is more involved. Unlike the MTPI case, $B_{\mathrm{fit,CW}}(q)$ is not flat (Fig. S2b) and is not representative of the true electronic offset/noise. Therefore, another approach was needed to estimate the electronic noise contribution in the CW DICF. For this, we work under the assumption that the electronic noise in CW is proportional to that in MTPI. Moreover, we observed that for both CW and MTPI, the DICF signal is constant (independent of either *q* and *τ*) for large q values (12–14 µm$^{-1}$, Fig. S3).

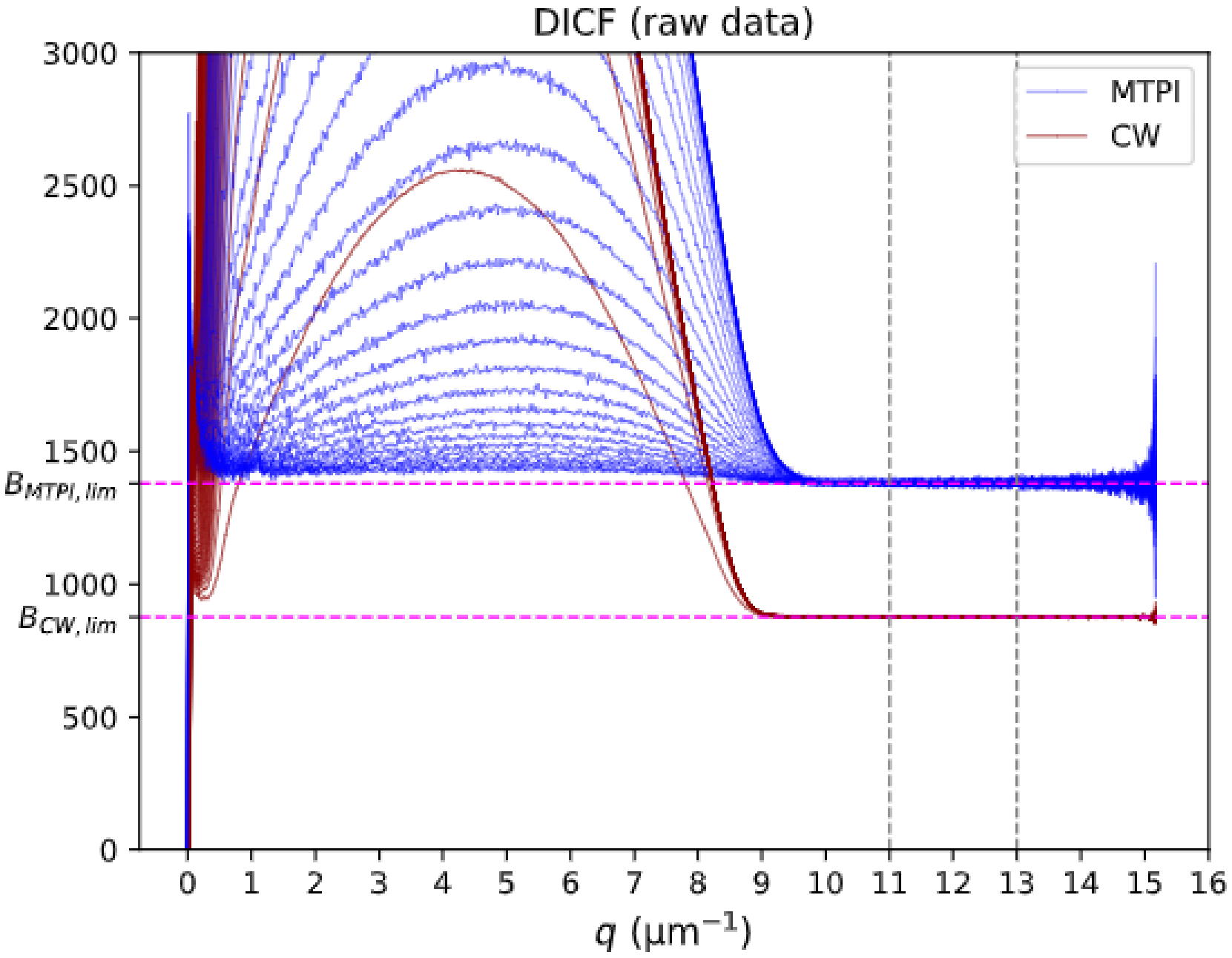

**Fig. S3.** Determination of background noise level $B_{\mathrm{MTPI,lim}}$ (resp. $B_{\mathrm{CW,lim}}$) from the azimuthally averaged raw DICF for pulsed (MTPI, blue) and CW (red) illumination as the constant value at high wavenumber.

*Figure description: Figure S3 shows the DICF curves as a function of q obtained using MTPI-DDM and CW illumination. The graph zooms in vertically on these curves to show that, from q = 10 µm$^{-1}$, the amplitude of the DICF no longer depends on q. This amplitude is higher in MTPI-DDM than that obtained in CW-DDM.*

For the CW (resp. MTPI) curve, we calculate $B_{\mathrm{CW,lim}}$ (resp. $B_{\mathrm{MTPI,lim}}$) as the mean value of the DICF over all *τ* and over *q* values between 11 and 13 µm$^{-1}$ (Fig. S3).

The electronic noise contribution to the CW DICF is then approximated as:

$$B_{\mathrm{CW}}(q) = \frac{B_{\mathrm{CW,lim}}}{B_{\mathrm{MTPI,lim}}} B_{\mathrm{fit,\,MTPI}}(q)$$

The corrected amplitude of the CW DICF is calculated as:

$$A_{\mathrm{CW}}(q) = A_{\mathrm{fit,CW}}(q) + B_{\mathrm{fit,CW}}(q) - B_{\mathrm{CW}}(q)$$

The CW-to-MTPI amplitude ratio $A_{\mathrm{CW}}(q)/A_{\mathrm{fit,MTPI}}(q)$ is plotted in Fig. S4.

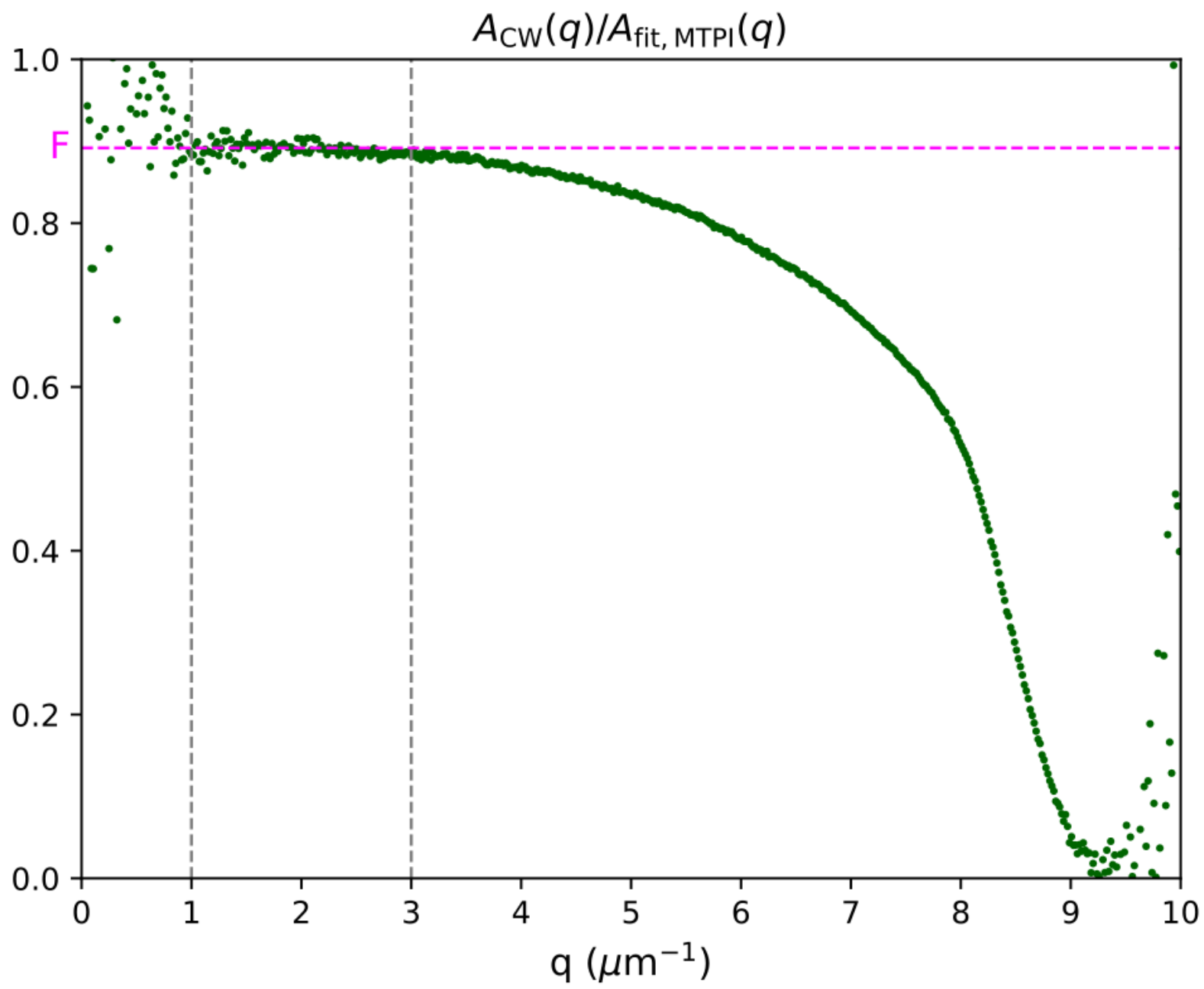

**Fig. S4.** Fit amplitude ratio $A_{\mathrm{CW}}(q)/A_{\mathrm{fit,MTPI}}(q)$ comparing continuous and pulsed illumination DDM.

*Figure description: Figure S4 shows a plot of the ratio between the amplitude of the DICF curves obtained in CW mode and that obtained in pulse mode, with the x-axis plotted against q for values ranging from 0 to 10 µm⁻¹. The curve exhibits a left-hand plateau with a height of approximately 0.9. On the portion of the curve corresponding to values of q between 1 and 3 µm⁻¹, the curve is practically horizontal. Beyond this range, the curve decreases.*

The amplitude ratio decreceases with *q* as a result of motion blur. An upper plateau at low *q* is observed because the blurring does not affect the DICF at low *q* values. We use the value of this plateau to rescale the CW DICF with respect to that of the MTPI. More precisely, we calculated the following rescaling factor:

$$F = \left\langle A_{CW}(q)/A_{fit,MTPI}(q) \right\rangle_{1<q<2}$$

The corrected CW DICF plotted in Fig. 3 and Fig. 4 of the Main Text were obtained using the following formula:

$$DICF_{CW,corrected}(q) = \big(DICF_{CW}(q) - B_{CW}(q)\big)/F$$

which enables a fair comparison between CW and MTPI modes.

The blurring factor, which is plotted Fig. 4c of the Main Text, is defined as:

$$\mathrm{Blur}(q) = (A_{CW}(q)/F)/A_{fit,MTPI}(q).$$